%% file: boundary-conditions_first_submission_to_ijss_and_arxiv.tex
\documentclass[10pt
]{article}
\usepackage[a4paper]{geometry}
\usepackage{graphicx}
\usepackage{amsmath,amssymb}
\usepackage{authblk}
\usepackage{cite}
\usepackage{subcaption}
\usepackage{xfrac}
\usepackage{authblk}
\usepackage{booktabs}
\usepackage{mathrsfs}

\usepackage{tikz}
\usetikzlibrary{decorations.pathmorphing}
\usetikzlibrary{decorations.markings}
\usetikzlibrary{shadows}
\usepackage{pgfplots}

\usepackage[abs]{overpic}

\usepackage[textsize=small,shadow]{todonotes} 

\input macro

\usepackage[defaultsans]{droidsans}
\usepackage[eulergreek]{sansmath}
\pgfplotsset{
  tick label style = {font=\sansmath\sffamily},
  every axis label = {font=\sansmath\sffamily},
  legend style = {font=\sansmath\sffamily},
  label style = {font=\sansmath\sffamily}
}

\newcommand{\vect}[1]{{\boldsymbol{#1}}}

\title
{Singular perturbations and cloaking illusions for elastic waves in membranes and  Kirchhoff  plates}

\author[1]{I.S. Jones}
\author[2]{M. Brun
}
\author[3]{N.V. Movchan}
\author[3]{A.B. Movchan}

\affil[1]{School of Engineering, John Moores University, Liverpool, L3 3AF, U.K.}
\affil[2]{Dipartimento di Ingegneria Meccanica, Chimica e dei Materiali, Universit\'{a} di Cagliari, Piazza d'Armi, I-09123 Cagliari, Italy}
\affil[3]{Department of Mathematical Sciences, University of Liverpool, Liverpool, L69 3BX, U.K.}

\begin{document}

\maketitle


\begin{abstract} A perturbation approach is used for analysis of a near-cloak in shielding a finite scatterer from an incident flexural wave.
The effect of the boundary conditions on the interior surface of the cloaking layer  is analysed in detail,
based on the explicit analytical solutions of a wave propagation problem for a membrane as well as a Kirchhoff flexural plate.
It is shown that the Dirichlet boundary condition on the interior contour of the cloak significantly reduces the cloaking action in the membrane case, and it also
makes cloaking impossible for flexural waves in a Kirchhoff plate.
\end{abstract}

\section{Introduction}

The cloaking of acoustic and electromagnetic waves has been extensively developed in terms of theoretical design and practical implementation in papers
\cite{PendrySchurigSmith2006,SchurigMockJusticeCummerPendrySmith2006,Leonhardt2006,Norris2008,ChenChan2010,GuenneauMcPhedranEnochMovchanFarhatNicorovici2011,KadicBuckmannSchittnyWegener2013,ZigoneanuPopaCummer2014}.
These publications  are based on the transformation optics approach, that may employ a non-conformal push-out map.
The theory equally applies to linear waterwave problems, as discussed in \cite{FarhatEnochGuenneauMovchan2008}.

Elastic cloaking, both for vector problems of elasticity and for flexural waves in Kirchhoff plates, brings new challenges regarding the physical interpretation of equations and boundary conditions in the cloaking region. Indeed, this is now well understood in the context of a lack of invariance of the governing equations of elasticity with respect to a non-conformal cloaking transform.  Analysis of elastic cloaking problems had been reported in \cite{MiltonBrianeWillis2006,BrunGuennauMovchan2009,NorrisShuvalov2011,NorrisParnell2012}.

In the recent papers \cite{FarhatGuenneauEnoch2009,FarhatGuenneauEnochMovchan2009,StengerWilhelmWegener2012,BrunColquittJonesMovchanMovchan2014,ColquittBrunGeiMovchanMovchanJones2014}, an advance has been made in the theoretical analysis, design and the physical interpretation of a cloak for flexural waves in Kirchhoff plates. Although it is known that membrane waves (solutions of the Helmholtz equation)  drive the wave propagation in Kirchhoff plates (see, for example, \cite{McPhedranMovchanMovchan2009,AntonakakisCraster2012,BrunGiaccuMovchanMovchan2012,CartaBrunMovchan2014,McPhedranMovchanMovchanBrunSmith2014}), the interface and boundary conditions provide coupling between solutions of the Helmholtz and modified Helmholtz equations in problems of scattering of flexural waves. In \cite{BrunColquittJonesMovchanMovchan2014,ColquittBrunGeiMovchanMovchanJones2014}, we have analysed in detail the transformed plate equation, and have shown that in the cloaking region the physical interpretation requires the presence of prestress and body force terms. Subject to such an allowance, the cloaking of defects in Kirchhoff plates becomes feasible and well understood. In the  present paper, we make an emphasis on the important issue of the choice of boundary conditions, which are set at the interior contour of the cloak.

In cloaking transformation problems, it is common that
boundary conditions on the interior boundary of a cloaking region are not addressed  
\cite{FarhatGuenneauEnoch2009,FarhatGuenneauEnochMovchan2009}.
This matter is rarely discussed and numerical simulations for cloaking are commonly
presented without additional comments regarding these boundary conditions.
The reason is that for a singular map, which ``stretches'' a hole of zero radius into a finite disk, in the unperturbed configuration there is no boundary.
Nevertheless, in the subsequent numerical computations the singularity of the material constants in the cloaking region is replaced by regularised finite values.
Namely, the boundary conditions are required for the numerical computations and usually natural boundary conditions that follow from the variational formulation are chosen, i.e. these are Neumann boundary conditions on the interior contour of the cloaking region \cite{NorrisParnell2012,ParnelNorrisShearer2012}.

One can ask a naive question related to an illusion rather than cloaking. For example, if one has a carrot cake, would it be possible to make it look like a fairy cake instead. In turn, could one make a large void in a solid look smaller? Purely naively, the latter can be addressed through a geometrical transformation in polar coordinates: 
\begin{equation*}
r=\alpha_1+\alpha_2 R,\quad \theta=\Theta,
\end{equation*}
where
\begin{equation*}
\alpha_1=\frac{R_2(R_1-a)}{R_2-a},\quad
\alpha_2=\frac{R_2-R_1}{R_2-a}.
\end{equation*}
and
\begin{equation*}
a \leq R \leq R_2, \quad R_1 \leq r \leq R_2.
\end{equation*}
with $a$ being a small positive number and $(R, \Theta)$ being the coordinates before the transformation and $(r, \theta)$ after the transformation, and $R_1, R_2$ being positive constants, such that  $R_1 < R_2$.
Such a transformation, within the ring $R_1 < r < R_2$,  would correspond to a radial non-uniform stretch, and assuming that outside the disk $r > R_2$ there is no deformation (i.e. $r=R$), we obtain a coating that would lead to an illusion regarding the size of a defect in a solid. Push-out transformations of such a type have been widely used, and have been applied to problems of cloaking, for example, in papers
\cite{Leonhardt2006,GreenleafLassasUhlmann2003}.

Formally, if $u$ represents an undistorted field in the exterior of a small void, of radius $a$ and we assume the series representation
$$
u(R, \Theta) = \sum_{n=-\infty}^\infty V_n(R) W_n(\Theta),
$$ 
with $V_n$ and $W_n$ being basis functions, then the distortion, represented by the fields $u_1$ and $u_2$ can be described through a ``shifted'' series representation, as follows
 $$
u= u_1(r, \theta) = \sum_{n=-\infty}^\infty V_n\left(\frac{r-\alpha_1}{\alpha_2}\right) W_n(\theta), ~ \mbox{when} ~  R_1 < r < R_2
$$ 
and 
$$
u= u_2(R, \Theta) = \sum_{n=-\infty}^\infty V_n(R) W_n(\Theta), ~ \mbox{when} ~ r > R_2.
$$ 
Such a transformation delivers an illusion,  which makes a finite circular void of radius $R_1$, look like a small void of radius $a$.

The above argument may appear to be simplistic, but we are going to show that it works exactly as described, for a class of problems governed by the Helmholtz operator and by the equations of vibrating Kirchhoff plates. In this cases, the basis functions $V_n$ and  $W_n$ are chosen accordingly, and are written in the closed form in the main text of the paper.

We consider the problem in the framework of singular perturbations and, instead of regularising the singular values of material parameters after the cloaking transformation, we begin by introducing a small hole and apply the cloaking transformation to a region containing such a small hole with appropriate boundary conditions already chosen. This is consistent with the analysis presented in \cite{KohnShenVogeliusWeinstein2008,ColquittJonesMovchaMovchaBrunMcPhedran2013}.

In the present paper, we show that the cloaking problem is closely related to that for an infinite body containing a small hole of radius $a$ where we set a boundary condition of either the Dirichlet (clamped boundaries) or Neumann (free-edge boundaries) type.
This is done for both membrane waves governed by the Helmholtz equation and flexural waves that occur in Kirchhoff plates.
Analytical solutions are presented for the cloaking problems and it is shown that the degree of cloaking is highly dependent on the boundary condition on the interior contour of the cloaking region.
In a plate, in the presence of an incident plane wave along the $x$-axis, the scattered flexural field, $u_s $, outside the cloaking region, when $a \to 0$ with a clamped interior boundary, has the asymptotic representation at a sufficiently large distance $R$ from the centre of the scatterer

\beq
u_s \sim - \sqrt{\fr{2}{\pi \beta  R}} \exp\left(i \left(\beta R - \fr{\pi}{4}\right)\right), ~~ \mbox{as}~ \beta R\to \infty,
\eequ{us1a}
where $\beta^4=\rho h \omega^2 /D_0$ with radian frequency $\omega$, plate flexural rigidity $D_0$, plate thickness $h$ and density $\rho$.
This immediately suggests the absence of any cloaking action as the above asymptotic representation corresponds to a finite point force initiated by a rigid pin at the origin (see \cite{EvansPorter2007,McPhedranMovchanMovchan2009}).
On the contrary,
the free-edge interior boundary in the cloaking layer for a flexural plate produces high-quality cloaking action, and the corresponding asymptotic representation of the scattered field $u_s$ at sufficiently large  $\beta R$ becomes

 \beq
u_s \sim  \sqrt{\fr{2}{\pi \beta  R}} \exp\left(i \left(\beta R - \fr{\pi}{4}\right)\right)  \frac{\pi i \nu }{4(1-\nu)} (\beta a)^2 = O(\beta  a)^2, ~~
\mbox{as }~~ \beta a \to 0 ~~\mbox{and} ~~  \beta R \to \infty.
\eequ{us2a}
This confirms that the regularisation algorithm that refers to a small free-edge hole produces a scattered field proportional to the area of this small hole, and it tends to zero as $\beta a \to 0.$

For the ``cloaking type" map considered in this paper, the original hole does not have to
be small, and we will refer to a ``cloaking illusion'', which results in an object being mimicked by an obstacle of a different size.

The structure of the paper is as follows. In Section \ref{Sec2} we introduce the notion of a cloaking transformation.
Section \ref{helmprob2a} addresses the singular perturbation approach for an elastic membrane. In particular, in Section  \ref{helmprob2}   we include an analytical solution, together with asymptotic estimates, for the model problem of scattering of membrane waves from a small circular scatterer.
Section \ref{Sec4}
shows the relationship between the model problem for a body with a small scatterer
and the full cloaked problem.
Section \ref{Sec5} presents the analytical solution and the asymptotic analysis of scattering of flexural waves in a Kirchhoff plate for the biharmonic cloaking problem. The concluding remarks are included in Section \ref{Sec6}, where we summarise the fundings regarding the cloaking action for different types of boundary conditions at the interior boundary of the cloaking region.

\section{Cloaking transformation}
\label{Sec2}

We aim to consider flexural waves in thin elastic plates. As shown in \cite{BrunColquittJonesMovchanMovchan2014,ColquittBrunGeiMovchanMovchanJones2014}, after the cloaking transformation such waves can be interpreted as time-harmonic flexural displacements  in a pre-stressed anisotropic thin plate. Firstly, we review the cloaking transformation  procedure and then we discuss the transformed equations and their physical interpretation.

\subsection{Push-out transformation}

A non-conformal transformation is introduced to define a cloaking coating.
We use the notations $\vect{X}=(R,\Theta)^T$ and $\vect{x}=(r,\theta)^T$ for coordinates before the transformation and after the transformation, respectively. Here $(R, \Theta)$ are polar coordinates.
With reference to \cite{PendrySchurigSmith2006,SchurigMockJusticeCummerPendrySmith2006,Leonhardt2006,GreenleafLassasUhlmann2003,Norris2008}, we use the radial invertible ``push-out'' map  $\mathcal{F}: ~\vect{X} \to \vect{x}$ defining the new stretched coordinates $(r, \Gt)$ as follows.
When
$|\vect{X}| < R_2$, the transformation $\vect{x} = \mathcal{F}(\vect{X})$ is given by 
\begin{equation}
            r=R_1+\dfrac{(R_2-R_1)}{R_2}  R,  
           \,\,\, \theta=\Theta,
\label{eqn001}
\end{equation}
so that $R_1<r<R_2$ when $0 < R < R_2$, with $R_1$ and $R_2$ being the interior and exterior radii of the cloaking ring, respectively.

In the exterior of the cloaking region, when $|\vect{X}| > R_2$, the transformation map is defined as the identity, so that $\vect{x} = \vect{X}$.

In the cloaking region, the Jacobi matrix ${\bf F} = D \mathcal{F} / D \vect{X}$
has the form
\begin{equation}
{\bf F}=
\frac{R_2-R_1}{R_2}{\bf e}_r\otimes{\bf e}_r+\frac{R_2-R_1}{R_2}\frac{r}{r-R_1}{\bf e}_\theta\otimes{\bf e}_\theta,
\label{eqn002}
\end{equation}
where ${\bf e}_r={\bf e}_R$, ${\bf e}_\theta={\bf e}_\Theta$ is the cylindrical orthonormal basis and $\otimes$ stands for the dyadic product.

After the above transformation, the classical Laplace's operator will be modified accordingly. For a given real $\alpha$, which is identified with $R_1$ for transformation (\ref{eqn001}), a new differential operator $\hat\nabla^2_{\alpha}$ is defined as
\begin{equation}
\hat\nabla^2_{\alpha}=\frac{1}{r-\alpha}\frac{\partial}{\partial r}\left[(r-\alpha)\frac{\partial}{\partial r}\right]+\frac{1}{(r-\alpha)^2}\frac{\partial^2}{\partial \theta^2}.
\label{eqn003}
\end{equation}
This is consistent with notations in the paper \cite{BrunColquittJonesMovchanMovchan2014}.
The operator $\hat\nabla^2_\alpha$ is referred to as the `shifted Laplace operator'.
As in \cite{BrunColquittJonesMovchanMovchan2014}, we also use the terms `shifted Helmholtz' and `shifted modified Helmholtz' for the operators $\hat\nabla^2_
\alpha + \beta^2$ and $\hat\nabla^2_
\alpha - \beta^2$, respectively.

\subsection{Physical interpretation of transformation cloaking for a membrane}

With the reference to the work of Norris \cite{Norris2008}, we give an outline of the transformed equation.
Firstly, the governing equation for a time-harmonic out-of-plane displacement $u$ of an elastic membrane has the form
\beq
\left(\nabla_{\vect{X}}\cdot\mu\nabla_{\vect{X}} + \rho\omega^2\right)u(\vect{X}) = 0,\; \vect{X}\in\mathbb{R}^2,
\eequ{eqn004}
where $\mu$,  $\rho$, and $\omega$ represent the stiffness matrix, the mass density and the radian frequency respectively.

On application of the transformation  $\vect{x} = \CF(\vect{X})$ within the cloaking region,  the transformed equation becomes
\beq
\left(\nabla\cdot\mu\vect{C}(\vect{x})\nabla + \frac{\rho\omega^2}{J(\vect{x})}\right)u(\vect{x}) = 0,
\eequ{eqn005}
where
\beq
\vect{C} = \frac{\vect{F}\vect{F}^\mathrm{T}}{J},\quad {\bf F} = \nabla_{\vect{X}}{\vect{x}},\quad J = \det \vect{F}.
\eequ{eqn006}
We note that equation \eq{eqn005}, similar to \eq{eqn004},  describes a vibrating membrane, but with different elastic stiffness, which is inhomogeneous and orthotropic, and a non-uniform distribution of mass across the transformed region.

In contrast, for the model of a flexural plate, it was shown in \cite{BrunColquittJonesMovchanMovchan2014,ColquittBrunGeiMovchanMovchanJones2014} that an additional pre-stress and body force are required to provide a full physical interpretation of the equations of motion in the transformed region, which again corresponds to an inhomogeneous and orthotropic material.
The fourth order governing equation for the plate is to be discussed in section \ref{Sec5}.
Special attention will be given to boundary conditions on the interior of the cloaking region.

Prior to the geometrical transformation,
there was no need to prescribe any boundary conditions at the origin.
On the other hand, after the transformation, one has an interior boundary of a finite size, and boundary conditions are required.
In particular, this issue always occurs if numerical simulations are carried out for an invisibility cloak using a finite element method.  On many occasions, such boundary conditions are chosen as natural boundary conditions and this means conditions of the Neumann type, which correspond to a free boundary at the interior of the cloaking layer.

The question arises as to whether or not imposing a Dirichlet condition on the interior boundary of the cloaking layer would make a difference in a numerical simulation or an experimental implementation of the cloak.  The answer to this question is affirmative, and it is linked to the analysis of a class of singularly perturbed problems introduced below.

\section{Singular perturbation problem in a membrane}
\label{helmprob2a}

It will be shown that a regularised cloak in a membrane
mimics a small circular scatterer and furthermore, if the radius of such a scatterer tends to zero the perturbation of the incident field due to an interaction with the obstacle reduces to zero as well. However, the asymptotic behaviour as the radius of the scatterer $a \to 0$, does depend on the type of boundary conditions on the scatterer.

\subsection{Model problem: scattering of a plane wave by a circular obstacle in a membrane}
\label{helmprob2}
Consider an unbounded homogeneous isotropic elastic solid with shear modulus $\mu$ and
density $\rho$ containing a
circular void of radius $a$.  Further, incident time-harmonic elastic out-of-plane shear wave of frequency $\omega$ results in a total scattered time-harmonic field of amplitude $u({\bf X})$
which obeys the Helmholtz equation outside the scatterer:

\beq
\left(\nabla^2 + k^2\right)u(\vect{X}) = 0,\;
\eequ{helm}
where $k=\omega \sqrt {\rho/\mu}$.

We choose polar coordinates $(R,\Theta)$ such that the direction of the incident plane waves is along the $\Theta=0$ line.  The total displacement field, obeying the radiation condition at infinity, may be written as

 \begin{equation}
u(R,\Theta)=\sum_{n=-\infty}^{\infty} \left[ i^n J_n(kR)- i^n p_nH_n^{(1)}(kR) \right] e^{in\Theta}
\label{helmprob1sol}
\end{equation}
where $J_n$ is the Bessel function of the first kind and $H_n^{(1)}$ is the first Hankel function. Use has also been made of the multipole expansion of the plane wave

\begin{equation}
e^{ikR\cos\Theta}=\sum_{n=-\infty}^{\infty}  i^n J_n(kR) e^{in\Theta}
\label{plane}
\end{equation}

The coefficients $p_n$ are determined by the boundary condition on the circular scatterer.    They are given by:
\begin{equation}
\label{alpha}
  p_n = \left\{
     \begin{array}{lr}
       \displaystyle{ \frac{J_n(ka)}{H_n^{(1)}(ka)}} &  : \text{if}  \quad  u=0 \quad \text{on} \quad R=a \\
       \\
       \displaystyle{ \frac{J_n'(ka)}{H_n^{(1)'}(ka)} }& \quad:   \text{if}  \quad  \dfrac{\prt u}{\prt R}=0 \quad \text{on} \quad R=a
\end{array}
\right.
\end{equation}
Thus the field may be written as
\begin{equation}
u(R,\Theta)=e^{ikR\cos \Theta}-p_0 H_0^{(1)}(kR)-\sum_{n=1}^{\infty}   i^n H_n^{(1)}(kR) (p_n e^{in\Theta}+p_{-n} e^{-in\Theta})
\label{genhelmprob1sol2}
\end{equation}
For the Dirichlet boundary condition $u=0$ on $R=a$, this may be re-written as
\begin{equation}
u(R,\Theta)=e^{ikR\cos \Theta}-\frac{J_0(ka)}{H_0^{(1)}(ka)} H_0^{(1)}(kR)-\sum_{n=1}^{\infty}  2 i^n\frac{J_n(ka)}{H_n^{(1)}(ka)} H_n^{(1)}(kR) \cos(n\Theta)
\label{helmprob1sol2}
\end{equation}

Correspondingly, for the case of the Neumann boundary condition $\prt u / \prt R =0$ on $R=a$, we have
\begin{equation}
u(R,\Theta)=e^{ikR\cos \Theta}-\frac{J_0'(ka)}{H_0^{(1)'}(ka)} H_0^{(1)}(kR)-\sum_{n=1}^{\infty}  2 i^n\frac{J_n'(ka)}{H_n^{(1)'}(ka)} H_n^{(1)}(kR) \cos(n\Theta)
\label{helmprob1sol3}
\end{equation}

In the asymptotic limit, when $ka \ll 1$ the coefficients near $H_0^{(1)}(kR)$ in the above formulae reduce to
\beq
-\frac{J_0(ka)}{H_0^{(1)}(ka)} \sim \fr{\pi i}{2 \log (k a)} \, \mbox{as} \ ka \to 0,
\eequ{as1H}
and
\beq
-\frac{J_0'(ka)}{H_0^{(1)'}(ka)} \sim -\fr{\pi i}{4} (ka)^2 \, \mbox{as} \ ka \to 0,
\eequ{as2H}
respectively.

We note that in both above cases, the coefficients near the monopole term tend to zero as $ka \to 0$. Furthermore, the coefficients near the higher-order multipole terms also vanish as $ka \to 0.$ With that said, the monopole coefficient in \eq{as1H}, corresponding to the Dirichlet problem, is of order $O(|\log (ka)|^{-1})$, i.e. decays logarithmically slowly.
Thus for a given size of small scatterer, the scattered field is reduced when Neumann boundary conditions are applied at the edge of the scatterer in comparison with the scattered field when Dirichlet boundary conditions are applied.

\subsection{Boundary conditions and the cloaking problem in a membrane}
\label{Sec4}

We note that the components of the stiffness matrix $\vect{C}$ in (\ref{eqn005}) after the ``cloaking transformation'' (\ref{eqn001}) are singular at the interior boundary of the cloaking layer.
In numerical computations, a regularisation is always introduced at the interior boundary to ``eliminate'' the singularity.

Alternatively, one can follow the approach advocated in \cite{ColquittJonesMovchaMovchaBrunMcPhedran2013} when a "near-cloak" is defined through the geometrical transformation of a plane with a small hole of radius $a$ into a plane containing a ring, with the unperturbed exterior radius $R_2$ and an
interior radius $R_1$.
Outside the ring, the transformation is equal to the identity.
Within the cloaking region, such transformation is defined by
\begin{equation}
\label{eqn007}
r=\alpha_1+\alpha_2 R,\quad \theta=\Theta,
\end{equation}
where
\begin{equation}
\label{eqn008}
\alpha_1=\frac{R_2(R_1-a)}{R_2-a},\quad
\alpha_2=\frac{R_2-R_1}{R_2-a}.
\end{equation}
and
\begin{equation}
\label{eqn009}
a \leq R \leq R_2, \quad R_1 \leq r \leq R_2.
\end{equation}

Outside the ring, the governing equation is \eq{eqn004}, whereas inside the ring the equation has the form \eq{eqn005}. For simplicity, assume that the material outside the ring is isotropic and homogeneous, i.e. $\mu = \mbox{const}$.
Then, equation \eq{eqn005} can also be written as
\begin{equation}
\label{eqn011}
\hat \nabla^2_{\alpha_1} u + \fr{\rho \omega^2}{\mu \alpha_2^2}  u = 0,
\end{equation}
where the operator $\hat \nabla^2_{\alpha_1}$ is defined by \eq{eqn003}.  It is also noted that $\alpha_1 \to R_1$ as $a \to 0.$

The domain with a small hole of radius $a$ has a singularly perturbed boundary in the limit when $a \to 0$. Simultaneously, this is also a regularisation of the push-out transformation used to design an invisibility cloak.

Thus, we have the following boundary value problem

\begin{equation}
\label{eqn012}
\left(\hat \nabla^2_{\alpha_1} + \fr{k^2}{\alpha_2^2} \right)u_1(r,\theta) = 0,\quad R_1 < r < R_2
\end{equation}

\beq
\left(\nabla^2 + k^2\right)u_2(R,\Theta) = 0,\; \quad R>R_2
\eequ{helm1}
where $k=\omega \sqrt {\rho/\mu}$.
The boundary conditions considered on the interior contour are
the Dirichlet boundary condition  (clamped boundary)
\begin{equation}
\label{bc1}
u_1(R_1,\theta) =0, \\[4 mm]
\end{equation}
or the Neumann boundary condition (free boundary)
\begin{equation}
\label{bc1a}
\frac{\prt u_1}{\prt r}(R_1,\theta) =0, \\[4 mm]
\end{equation}
together with transmission conditions at the interface $r = R_2$
\begin{equation}
\label{bc2}
u_1(R_2,\theta) =u_2(R_2,\Theta), \\[3 mm]
\end{equation}
\begin{equation}
\label{bc3}
\alpha_2 \frac{\partial u_1(R_2,\theta)}{\partial r} =
\frac{\partial u_1(R_2,\theta)}{\partial R} =
\frac{\partial u_2(R_2,\theta)}{ \partial R},
\end{equation}
and the radiation condition as $R \to \infty$.

The solution of the Dirichlet problem \eq{eqn011}, \eq{eqn012}, \eq{helm1}, \eq{bc1}, \eq{bc2}, \eq{bc3}  may be written as

\begin{equation}
\label{u1sol}
u_1(r,\theta)=\sum_{n=-\infty}^{\infty} \left[ i^n J_n\left(\frac{k}{\alpha_2}(r-\alpha_1)\right)-i^n\frac{J_n(ka)}{H_n^{(1)}(ka)} H_n^{(1)}\left(\frac{k}{\alpha_2}(r-\alpha_1) \right) \right]e^{in\theta}
\end{equation}
and
\begin{equation}
\label{u2sol}
u_2(R,\Theta)=\sum_{n=-\infty}^{\infty} \left[ i^n J_n(kR)-i^n\frac{J_n(ka)}{H_n^{(1)}(ka)} H_n^{(1)}(kR) \right]e^{in\Theta}
\end{equation}

It is important to note that the outer solution $u_2(r,\theta)$ is the field of particular interest when examining the quality of cloaking. It is {\it exactly the same} as the field produced by a circular scatterer of radius $a$
with Dirichlet boundary condition on the scatterer (equations (\ref{helmprob1sol}) and (\ref{alpha})). Thus the degree of disturbance of the incident plane wave
is totally determined by the effect
of this single uncloaked scatterer. In turn, this is then dependent on both the size of the circular scatterer and the boundary conditions on it.

The formulae \eq{u1sol} and \eq{u2sol} give the closed form analytical solution for a regularised invisibility cloak subject to the incident plane wave and the Dirichlet boundary condition on the interior contour of the cloaking layer.

In a similar way, when the boundary condition on the interior contour is the Neumann boundary condition \eq{bc1a}, the formulae \eq{u1sol} and \eq{u2sol} are replaced by
\begin{equation}
\label{u1solne}
u_1(r,\theta)=
\sum_{n=-\infty}^{\infty} \left[ i^n J_n\left(\frac{k}{\alpha_2}(r-\alpha_1)\right)-i^n \frac{J_n'(ka)}{{H_n^{(1)}}'
(ka)}
H_n^{(1)}\left(\frac{k}{\alpha_2}(r-\alpha_1) \right) \right]e^{in\theta}
\end{equation}
and
\begin{equation}
\label{u2solne}
u_2(R,\Theta)=\sum_{n=-\infty}^{\infty} \left[ i^n J_n(kR)-i^n\frac{J_n'(ka)}{{H_n^{(1)}}'(ka)} H_n^{(1)}(kR) \right]e^{in\Theta}.
\end{equation}

\begin{figure}[!htcb]
\centerline{
         \includegraphics[width=.9\columnwidth]{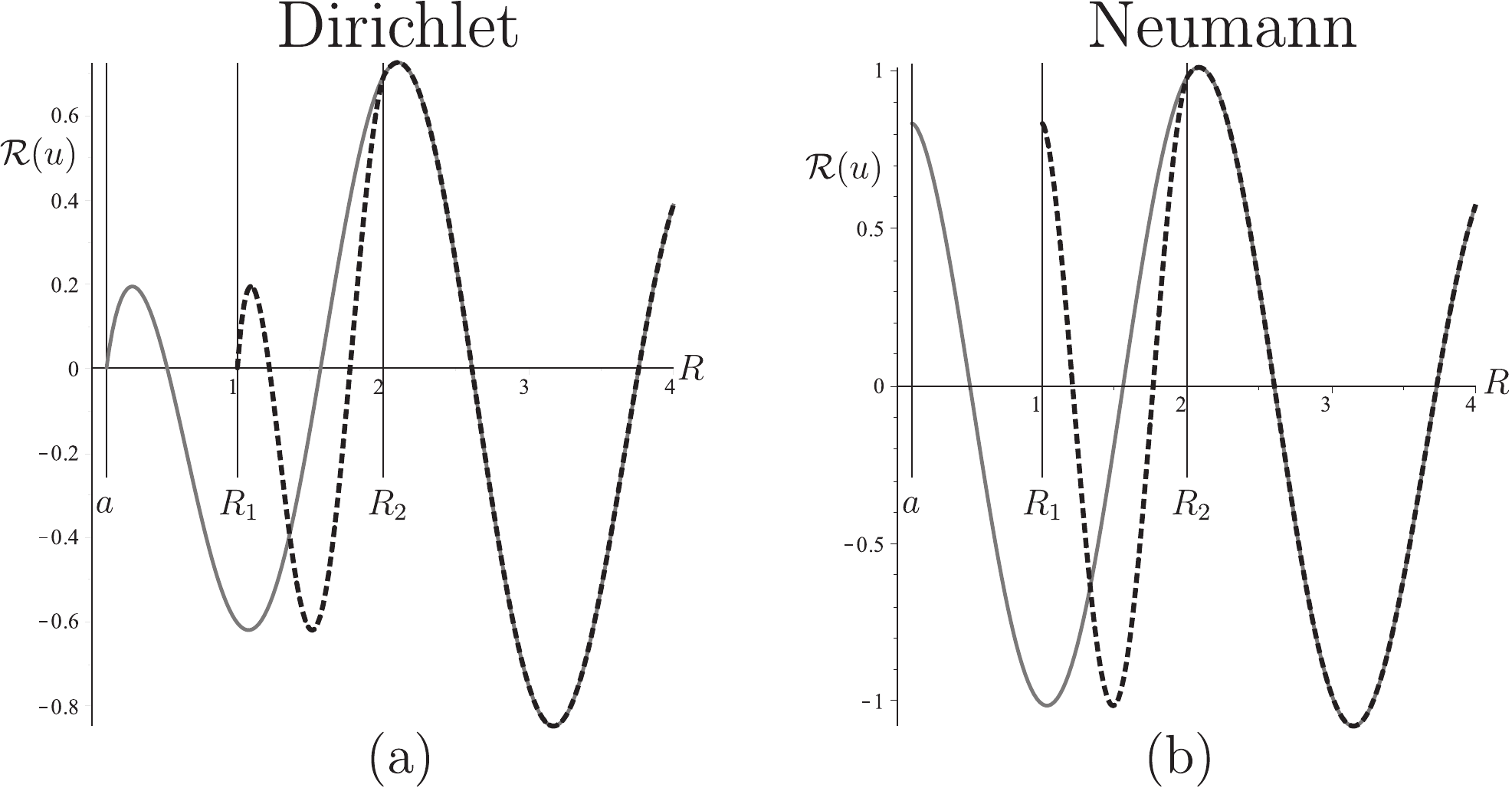}
}
\caption{
Membrane Problem. Displacement $\mathcal{R}[u(R,0)]$ in the direction of the incident plane-wave propagation as a function of the distance $R$ from the center of the scatterer or the cloaked region. Curves are given for the parameters $a=0.1$ cm, $R_1=1$ cm, $R_2=2$ cm, corresponding to $\alpha_1=0.947$ and $\alpha_2=0.526$ cm, and $k=3$ cm$^{-1}$.
(a) Dirichlet problem: displacement in a homogeneous membrane with circular obstacle (grey continuous line, equation (\ref{genhelmprob1sol2})) and in the cloaking problem (black dashed line, equations (\ref{u1sol},\ref{u2sol})). (b) Neumann problem: displacement in a homogeneous membrane with circular obstacle (grey continuous line, equation (\ref{helmprob1sol2})) and in the cloaking problem (black dashed line, equations (\ref{u1solne},\ref{u2solne})). 
The sums in the multipole expansion representations of the displacements are truncated after the first $10$ terms.}
\label{Fig01}
\end{figure}

The exterior field outside the cloaking layer is exactly the same as for the case of a small circular scatterer in section \ref{helmprob2} with either Dirichlet or Neumann boundary conditions. The displacement fields for the homogeneous membrane with a circular scatterer and the cloaking problem are shown in Figure \ref{Fig01}; they are given for the two types of boundary conditions. As  $a\to0$ and according to \eq{as1H} and \eq{as2H} the quality of cloaking due to the cloaking layer with the Neumann boundary condition is higher than that due to the cloaking layer with the Dirichlet boundary condition. This is because the coefficient near the leading order in the scattered component of the solution is $O(|\log (ka)|^{-1})$ for the Dirichlet case and $O((ka)^2)$ for the Neumann case.

\section{Singular perturbation and cloaking action for the biharmonic problem}
\label{Sec5}

Although it appears that both Dirichlet and Neumann boundary conditions for the membrane can be used for the design of the invisibility cloak, we have demonstrated that the logarithmic asymptotics in \eq{as1H} and the power asymptotics of \eq{as2H} suggest that the Neumann condition produces a more efficient cloak compared to the Dirichlet cloak as  $a\to0$.

This message is further reinforced by considering flexural waves.
We will now consider the analytical solution to a further model problem when a  plane flexural wave in a Kirchhoff plate is incident on a small circular scatterer together with a corresponding cloaking problem in an analogous manner to that of sections \ref{helmprob2} and \ref{Sec4}. Below, we will consider the biharmonic operator instead of the Helmholtz operator.

\subsection{A model problem of scattering of a flexural wave by a circular scatterer}
\label{flex_asymp}

Consider an unbounded homogeneous isotropic Kirchoff plate with flexural rigidity $D_0$, thickness $h$ and density $\rho$ containing a clamped-edge hole of radius $a$.  Further, incident time-harmonic elastic flexural plane waves of frequency $\omega$ result in a total scattered time-harmonic field of amplitude $u({\bf X})$, at position ${\bf X}$. As in section \ref{helmprob2}, cylindrical coordinates  $(R,\Theta)$ will be used such that the direction of the incident plane waves is along the $\Theta=0$ line. The field $u(R,\Theta)$ obeys the following problem outside the scatterer which is regarded as clamped,

\beq
(\Delta^2 - \beta^4 ) u(R,\Theta) = 0,\quad R > a;~~
\eequ{prob3a}
\beq
u=0, \quad \dfrac{\partial u}{\partial R}=0, \quad R=a,
\eequ{prob3}
where $\beta^4=\rho h \omega^2 /D_0$.

The problem is solved by factorising the operator into Helmholtz and modified Helmholtz operators in the usual way. Applying the radiation condition at infinity leads to the total field being composed of three contributions from the original plane wave, the Helmholtz  and the modified Helmholtz fields. The general solution is

 \begin{equation}
 \label{bihgensol}
u(R,\Theta)=\sum_{n=-\infty}^{\infty} \left[ i^n J_n(\beta R)+p_nH_n^{(1)}(\beta R)+q_nK_n(\beta R) \right] e^{in\Theta}
\end{equation}
where $K_n$ is the modified Bessel function. Use has also been made of the plane wave expansion in cylindrical coordinates given in equation (\ref{plane}). Application of the clamped boundary conditions in equation (\ref{prob3}) leads to the following expressions for the coefficients $p_n$ and $q_n$

\begin{equation}
\label{rands}
\begin{bmatrix} p_n  \\q_n \end{bmatrix} =
\dfrac{1}{\CW[H_n^{(1)}(\beta a),K_n(\beta a)]}
\begin{bmatrix} K_n'(\beta a) & -K_n(\beta a)  \\-{H_n^{(1)}}'(\beta a) & H_n^{(1)}(\beta a) \end{bmatrix}
\begin{bmatrix} -i^n J_n(\beta a) \\-i^n J_n'(\beta a) \end{bmatrix}
\end{equation}
where
\begin{equation}
\nonumber
\CW[H_n^{(1)}(\beta a),K_n(\beta a)]=H_n^{(1)}(\beta a)K_n'(\beta a)-{H_n^{(1)}}'(\beta a)K_n(\beta a)
\end{equation}

For the case of free-edge boundary, the normal component of the moment and the transverse force are equal to zero on the boundary of the small circular scatterer, which results in the following conditions to be satisfied

\begin{equation}
\left[\frac{\partial^2 u}{\partial R^2}+\nu\left(\frac{1}{R}\frac{\partial u}{\partial R}+\frac{1}{R^2}\frac{\partial^2 u}{\partial \Theta^2}\right)\right]=0, ~~\mbox{at} ~~ R = a,
\label{mom}
\end{equation}
and
\begin{equation}
T_R(u) = 0, ~~\mbox{at} ~~ R = a,
\label{Tmodel}
\end{equation}
where
\begin{equation}
T_R(u)=-\left[\frac{\partial}{\partial R}\left(\frac{\partial^2 u}{\partial R^2 }+\frac{1}{R}\frac{\partial u}{\partial R}+\frac{1}{R^2}\frac{\partial^2 u}{\partial \Theta^2}\right)\right]-\frac{1}{R}\frac{\partial M_{RT}(u)}{\partial \Theta}
\label{bigT}
\end{equation}
with
\begin{equation}
M_{RT}(u)=(1-\nu)\left[\frac{1}{R}\frac{\partial^2 u}{\partial R \partial \Theta}-\frac{1}{R^2}\frac{\partial u}{\partial \Theta}\right]
\label{bigMRT}
\end{equation}


Application of the free boundary conditions in equation (\ref{prob3}) instead of the clamped boundary conditions leads to the following expressions for the coefficients $p_n$ and $q_n$
\begin{equation}
\label{randsfree}
\begin{bmatrix} p_n  \\q_n \end{bmatrix} =
\dfrac{1}{{\mathscr D}}
\begin{bmatrix} V^{-}(K_n(\beta a)) & -X^{+}(K_n(\beta a))  \\-V^{+}(H_n^{(1)}(\beta a)) & X^{-}(H_n^{(1)}(\beta a)) \end{bmatrix}
\begin{bmatrix} -i^n  X^{-}(J_n(\beta a)) \\-i^n V^{+}(J_n(\beta a) )\end{bmatrix}
\end{equation}
where
\begin{equation}
X^{\pm}({\mathscr C}_n(\beta a))=-\beta a(1-\nu){\mathscr C}_n'(\beta a)+[n^2(1-\nu)\pm (\beta a)^2]{\mathscr C}_n(\beta a),
\end{equation}
\begin{equation}
V^{\pm}({\mathscr C}_n(\beta a))=[\pm (\beta a)^3+(1-\nu)n^2\beta a]{\mathscr C}_n'(\beta a)-(1-\nu)n^2{\mathscr C}_n(\beta a),
\end{equation}
with ${\mathscr C}_n(\beta a)$ denoting $H_n^{(1)}(\beta a)$ or $K_n(\beta a)$,
and
\begin{equation}
{\mathscr D}=X^{-}(H_n^{(1)}(\beta a))V^{-}(K_n(\beta a))-V^{+}(H_n^{(1)}(\beta a))X^{+}(K_n(\beta a)).
\label{D}
\end{equation}

The singular perturbation approach advocated above and the regularised push-out transformation work equally well for the case of a cloak re-routing flexural waves. Compared to the membrane waves, additional action is produced by evanescent waves; in particular, Green's function for flexural waves remains bounded whereas for the membrane problem it is singular. 

We refer to the paper \cite{Konenkov1964} that has addressed the classical solution of the scattering of a flexural plane wave.
Here we summarise the results, with the emphasis on the asymptotics of outgoing waves when $\beta a$ becomes small, $0< \beta a \ll 1$.
As in \cite{Konenkov1964}, the coefficient $p_0$ in the expansion  \eq{bihgensol} of the flexural displacement has the following limit representation as $\beta a \ll 1$
\beq
p_0 \sim -1, ~~ \mbox{as} ~~ \beta a \to 0,
\eequ{r0D}
for the case of the clamped boundary (Dirichlet problem), and
\beq
p_0 \sim \frac{\pi i \nu}{4(1-\nu)} (\beta  a)^2, ~~ \mbox{as} ~~ \beta a \to 0,
\eequ{r0N}
for the case of the free-edge boundary (Neumann problem). Furthermore, the coefficients $p_n, |n| \geq 1,$ vanish as $\beta a \to 0$, for both types of boundary conditions considered above.
We also note that the coefficients $q_n$ near $K_n$ terms are not significant, as these terms are exponentially small away from the scatterer, as $\beta R \gg 1.$

For the Dirichlet problem, when the contour of the circular inclusion is clamped as in \eq{prob3}, the scattered field  $u_s$ is
\beq
u_s \sim - \sqrt{\fr{2}{\pi \beta  R}} \exp\left(i \left(\beta R - \fr{\pi}{4}\right)\right) ~~ \mbox{as} ~~ \beta R \to \infty.
\eequ{us1}
For the case of a free-edge boundary condition, where both the transverse force and the moment vanish at $R = a$, the scattered field $u_s$ becomes
\beq
u_s \sim  \sqrt{\fr{2}{\pi \beta  R}} \exp\left(i \left(\beta R - \fr{\pi}{4}\right)\right)  \frac{\pi i \nu }{4(1-\nu)} (\beta a)^2 ~~ \mbox{as} ~~ \beta R \to \infty.
\eequ{us2}
Equation \eq{us1} clearly illustrates that, even in the limit of $\beta a \to 0$, the scattered field produced by the ``rigid  pin'' corresponds to a finite force. In other words, for the Dirichlet boundary condition, the singular perturbation of the boundary produces a finite scattered field and this remains finite in the neighbourhood of the scatterer even when the diameter of this scatterer tends to zero.
On the contrary, for the case of the free-edge boundary, as shown by \eq{us2}, the coefficient in the leading order term is $O((\beta a)^2)$, and it vanishes as $\beta a \to 0$.

Comparison with
\eq{as1H} and \eq{as2H} shows that, while for the Helmholtz operator, the monopole term in the scattered field always vanishes in the limit of $\beta a \to 0$ (both for Dirichlet and Neumann  boundary conditions); in the biharmonic case, which corresponds to a Kirchhoff plate, the scatterer with the clamped boundary (Dirichlet boundary condition) delivers a finite non-zero point force in the limit of $\beta a \to 0$. On the other hand, the case of a free-edge boundary (Neumann boundary condition) gives no scattering in the limit of   $\beta a \to 0$, as seen from \eq{us2}.

\subsection{Boundary conditions and the cloaking problem in a Kirchhoff plate}

The problem to be solved in this sub-section is analogous to that in section \ref{Sec4} except that we will consider cloaking of flexural rather than out-of-plane shear waves. A ``near-cloak" is again defined through the invertible geometrical transformation of a plate with a small hole of radius $a$, into a plate containing a ring, with unperturbed exterior radius $R_2$ and an expanded interior radius $R_1$ as in equations  (\ref{eqn007}), (\ref{eqn008}) and  (\ref{eqn009}).

Again, incident time-harmonic flexural waves of frequency $\omega$ result in a total scattered time-harmonic fields of amplitude $u_1(r,\theta)$ and $u_2(R,\Theta)$  inside and outside the cloak respectively.
The material outside the ring is isotropic and homogeneous and the governing equation is as in  Eq. (\ref{prob3a}).
Inside the cloak, the governing equation is given by the transformed biharmonic equation and the classical Laplace's operator is modified accordingly as in equation (\ref{eqn003}). The problem to be solved is defined as

\begin{equation}
\label{biprob4}
\left(\hat \nabla^4_{\alpha_1} - \fr{\beta^4}{\alpha_2^4} \right)u_1(r,\theta) = 0,\quad R_1 \leq r \leq R_2
\end{equation}

\beq
\left(\nabla^4 - \beta^4\right)u_2(R,\Theta) = 0,\; \quad R>R_2
\eequ{biprob42}

For the case of the clamped interior boundary, the boundary conditions are
\begin{eqnarray}
\label{bibc1}
u_1(R_1,\theta) =0, \\
\frac{\partial u_1(R_1,\theta)}{\partial r} =0.
\end{eqnarray}

For the case of free-edge boundary, zero transverse force and normal component of the moment on the interior contour of the cloaking region correspond to transformed boundary operators, obtained from the variational formulation for the shifted biharmonic operator in \eq{biprob4} used within the cloak. They result in the following conditions to be satisfied at $r=R_1$
\begin{equation}
{\mathfrak T}_r=-\alpha_2^3 \left[\frac{\partial}{\partial r}\left(\frac{\partial^2 u_1}{\partial r^2 }+\frac{1}{(r-\alpha_1)}\frac{\partial u_1}{\partial r}+\frac{1}{(r-\alpha_1)^2}\frac{\partial^2 u_1}{\partial \theta^2}\right)\right]-\frac{\alpha_2}{(r-\alpha_1)}\frac{\partial {\frak M}_{rt}}{\partial \theta}=0,
\end{equation}
with
\begin{equation}
{\frak M}_{rt}=\alpha_2^2(1-\nu)\left[\frac{1}{(r-\alpha_1)}\frac{\partial^2 u_1}{\partial r \partial \theta}-\frac{1}{(r-\alpha_1)^2}\frac{\partial u_1}{\partial \theta}\right],
\end{equation}
and
\begin{equation}
\alpha_2^2\left[\frac{\partial^2 u_1}{\partial r^2}+\nu\left(\frac{1}{(r-\alpha_1)}\frac{\partial u_1}{\partial r}+\frac{1}{(r-\alpha_1)^2}\frac{\partial^2 u_1}{\partial \theta^2}\right)\right]=0.
\end{equation}
In addition, on the interface between the cloak and the ambient material we have
\begin{eqnarray}
\label{bibc2}
u_1(R_2,\theta) =u_2(R_2,\Theta), \\
\alpha_2 \frac{\partial u_1(R_2,\theta)}{\partial r}
=\frac{\partial u_1(R_2,\theta)}{\partial R}
= \frac{\partial u_2(R_2,\Theta)}{\partial R},
\end{eqnarray}\begin{equation}
\left[\frac{\partial^2 u_2}{\partial R^2}+\nu\left(\frac{1}{R}\frac{\partial u_2}{\partial R}+\frac{1}{R^2}\frac{\partial^2 u_2}{\partial \Theta^2}\right)\right]=\alpha_2^2\left[\frac{\partial^2 u_1}{\partial r^2}+\nu\left(\frac{1}{(r-\alpha_1)}\frac{\partial u_1}{\partial r}+\frac{1}{(r-\alpha_1)^2}\frac{\partial^2 u_1}{\partial \theta^2}\right)\right],
\label{eq058a}
\end{equation}
for the normalised normal component of the moment
on the outer boundary of the cloak, and
\begin{equation}
T_R(u_2) = {\mathfrak T}_r(u_1)
\label{eq058b}
\end{equation}
for the normalised transverse force
on the outer boundary of the cloak. The transverse force $T_R(u)$ is defined in \eq{bigT} and \eq{bigMRT}.


We also supply  the radiation condition as $R \to \infty$, i.e. the scattered field is represented as an outgoing wave.

For the homogeneous material, when $\alpha_2 = 1$ and $\alpha_1=0$, the above conditions imply the continuity of the flexural displacement and rotation, in addition to the normal moments and transverse forces.

The series representation of the flexural displacement field inside the cloak and outside the cloaking region has the form
\begin{eqnarray}
\label{biu1sol}
\nonumber
u_1(r,\theta)=\sum_{n=-\infty}^{\infty} \left[ a_n J_n\left(\frac{\beta}{\alpha_2}(r-\alpha_1)\right)+b_n H_n^{(1)}\left(\frac{\beta }{\alpha_2}(r-\alpha_1) \right) \right.\\
\left. {+c_n I_n}\left(\frac{\beta }{\alpha_2}(r-\alpha_1) \right)+d_n K_n\left(\frac{\beta }{\alpha_2}(r-\alpha_1) \right) \right]e^{in\theta},
\,\,\, R_1< r < R_2,
\end{eqnarray}
and
\begin{equation}
\label{biu2sol}
u_2(R,\Theta)=\sum_{n=-\infty}^{\infty} \left[ i^n J_n\left(\beta R\right)+p_n H_n^{(1)}\left(\beta R \right)+q_n K_n\left(\beta R \right) \right]e^{in\Theta},
\,\,\,
R>R_2,
\end{equation}
respectively.

For each order $n$, there are six unknown coefficients.
For the case of the clamped interior boundary $r=R_1$, a solution satisfying the six boundary and interface conditions (\ref{bibc1}) - (\ref{eq058b}) is given by $a_n=i^n$, $b_n=p_n$, $c_n=0$,  $d_n=q_n$ together with $p_n$ and $q_n$ satisfying

\begin{equation}
\label{eandf}
\begin{bmatrix} p_n  \\q_n \end{bmatrix} =
\dfrac{1}{\CW[H_n^{(1)}(\beta a),K_n(\beta a)]}
\begin{bmatrix} K_n'(\beta a) & -K_n(\beta a)  \\-{H_n^{(1)}}'(\beta a) & H_n^{(1)}(\beta a) \end{bmatrix}
\begin{bmatrix} -i^n J_n(\beta a) \\-i^n J_n'(\beta a) \end{bmatrix},
\end{equation}
where
\begin{equation}
\nonumber
\CW[H_n^{(1)}(\beta a),K_n(\beta a)]=H_n^{(1)}(\beta a)K_n'(\beta a)-{H_n^{(1)}}'(\beta a)K_n(\beta a).
\end{equation}

Note that these equations for  $p_n$ and $q_n$ are the same as 
in \eq{rands} for  the case of the clamped boundary $R=a$, and the corresponding coefficients for the case of the free-edge condition are the same as  $p_n$ and $q_n$  in  equation \eq{randsfree}, respectively.

\begin{figure}[!htcb]
\centerline{
         \includegraphics[width=.9\columnwidth]{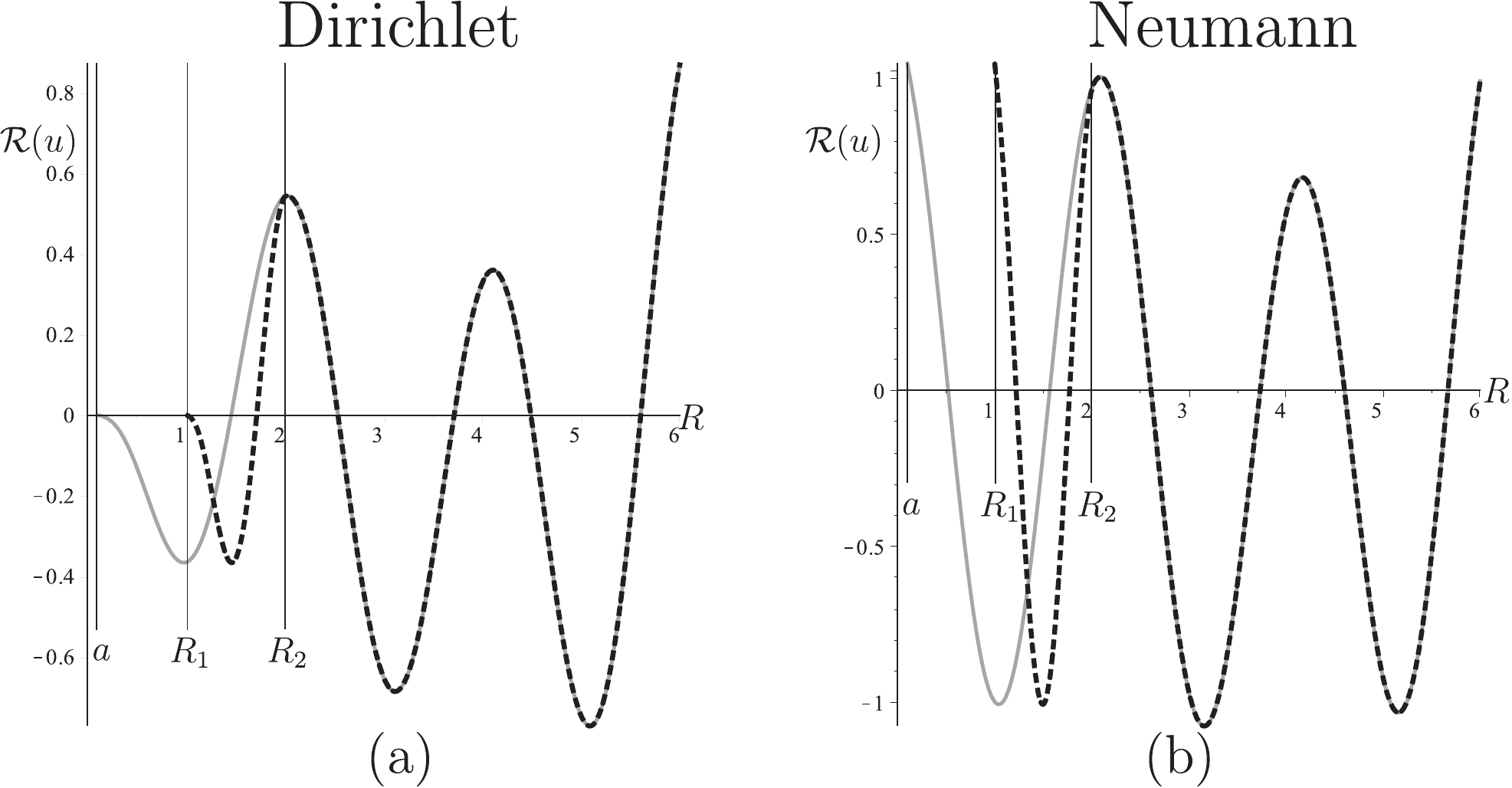}
}
\caption{
Plate Problem. Displacement $\mathcal{R}[u(R,0)]$ in the direction of the incident plane-wave propagation as a function of the distance $R$ from the center of the scatterer or the cloaked region. Curves are given for the parameters $a=0.1$ cm, $R_1=1$ cm, $R_2=2$ cm, corresponding to $\alpha_1=0.947$ and $\alpha_2=0.526$ cm, $\beta=3$ cm$^{-1}$ and $\nu=0.3$.
(a) Dirichlet problem: displacement in a homogeneous plate with circular obstacle (grey continuous line, equations (\ref{bihgensol},\ref{rands})) and in the cloaking problem (black dashed line, equations  (\ref{biu1sol}-\ref{eandf})). (b) Neumann problem: displacement in a homogeneous plate with circular obstacle (grey continuous line, equations (\ref{bihgensol},\ref{randsfree})) and in the cloaking problem (black dashed line). 
The sums in the multipole expansion representations of the displacements are truncated after the first $10$ terms.}
\label{Fig02}
\end{figure}

This implies that the outer field $u_2(R,\Theta)$ for the cloaking problem is {\it exactly the same} as the field produced by
a circular scatterer subjected to the same boundary conditions (either clamped or free-edge boundary) as those on the interior boundary of the cloak in the cloaking problem.
The required solutions are given by equations (\ref{bihgensol})--(\ref{D}). The displacement fields are shown in Figure \ref{Fig02}.

The asymptotic representation of the displacement field $u_2$,  for $R \gg 1$, outside the regularised cloak, considered here, is also
the same as the one for the small inclusion with either Dirichlet or Neumann boundary conditions, as in section
\ref{flex_asymp}. In particular, according to the formulae \eq{us1} and \eq{us2}, the cloaking action is non-existent if the interior contour of the cloaking layer in the Kirchhoff plate is clamped  (Dirichlet boundary condition set on $r=R_1$). On the other hand, the cloaking appears to be efficient for the free-edge boundary,  with the leading term of the scattered field to be of order $(\beta a)^2$ as $\beta a \to 0$.

\section{Concluding remarks. Cloaking illusion for the membrane and the Kirchhoff plate problems}
\label{Sec6}

This paper dispels  a common perception of an ``invisibility'' based on a singular geometric transformation used in the design of invisibility cloaks.
It has been shown that the boundary conditions on the interior contour of the ``cloaking'' may influence significantly the scattered field.
Singular perturbation analysis is the tool which enables one to see it.

Based on the model of a near-cloak, for flexural waves in a Kirchhoff plate we have shown  that the object surrounded by the cloak, appears as an infinitesimally small scatterer, with appropriate boundary conditions on its contour. The intuitive perception of a scattered field that vanishes when the diameter of such a scatterer tends to zero proves to be wrong for the case of flexural waves scattered by a clamped small inclusion.

Even when a singular transformation is used to design an ``exact'' cloak, in the numerical implementation, or in a physical experiment, it is always regularised.
As noted earlier, such regularisation is linked to setting  boundary conditions on the interior contour of the cloaking region.

For the near-cloak, the
singular perturbation model shows that even for the case of an elastic membrane, governed by the Helmholtz equation, setting the homogenous  Dirichlet boundary conditions
would lead to a higher scattering compared to the case of homogeneous Neumann boundary conditions on the interior contour of the cloaking layer.
This observation is fully supported by the asymptotic formulae  \eq{as1H}, \eq{as2H} for the coefficients of the leading term of the scattered field for the near-cloak.
The coefficient $p_0$ near $H_0^{(1)}(k r)$ in the representation for the scattered field in the membrane, as $ka \to 0$, is
\begin{eqnarray}
\nonumber
p_0 \sim \fr{\pi i}{2 \log (k a)}, \quad \mbox{(Dirichlet problem)}, \\
p_0 \sim -\fr{\pi i}{4} (ka)^2, \quad \mbox{(Neumann problem)}.
\label{eqn101}
\end{eqnarray}

\begin{figure}[!htcb]
\centerline{
         \includegraphics[width=.9\columnwidth]{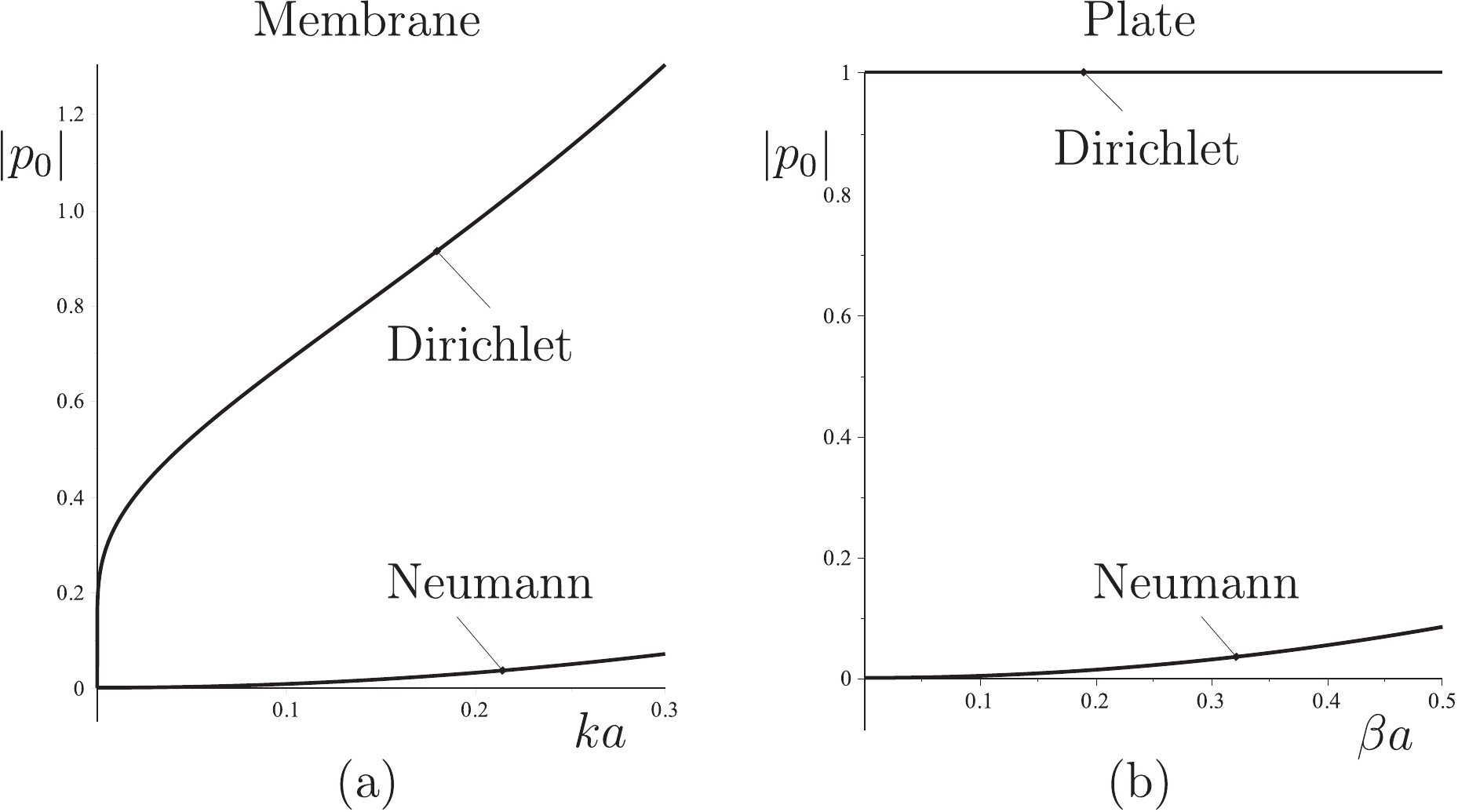}
}
\caption{
Monopole coefficient: modulus of $|p_0|$. 
(a) Membrane problem: $|p_0|$ is given for the Dirichlet and Neumann problems, equations (\ref{eqn101}).
(b) Plate problem: $|p_0|$ is given for the Dirichlet and Neumann problems, equations (\ref{eqn103}).}
\label{Fig03}
\end{figure}

Furthermore, the
message regarding the choice of the Neumann boundary conditions is reinforced by the analysis of the flexural waves re-routed by a cloak in a Kirchhoff plate. For the biharmonic problem describing such flexural waves,
we have shown that setting the full clamping (homogeneous Dirichlet boundary conditions) of the interior contour  would remove the cloaking action, even though the physical parameters of the cloak are  chosen according to the push-out transformation \eq{eqn001}.
On the other hand, the homogenous Neumann boundary conditions, corresponding to a free-edge interior contour of the cloak,
would be appropriate to the cloaking action
and hence routing membrane and flexural waves around a finite scatterer. For the Kirchhoff plate waves, the coefficient $p_0$ near $H_0^{(1)}(\beta r)$ in the representation for the scattered field is characterised by the following asymptotics as $\beta a \to 0:$
\begin{eqnarray}
\nonumber
p_0 \sim -1, \quad \mbox{(Dirichlet problem)}, \\
p_0 \sim \frac{\pi i \nu}{4(1-\nu)} (\beta  a)^2,\quad \mbox{(Neumann problem)}.
\label{eqn103}
\end{eqnarray}
The dependance of the monopole coefficient $p_0$ as a function of the small parameters $ka$ and $\beta a$ for the membrane and plate problems, respectively, is also shown in Figure (\ref{Fig03}).

Practical applications of filters, polarisers and cloaks for elastic flexural waves are in the efficient design of the earthquake protection systems.
The present paper shows that a choice of the type of boundary conditions on the interior contour of the cloak is essential.
The results of the paper  also have a wide range of  applications  concerning practical designs of cloaking shields for electro-magnetic and acoustic waves.

\section*{Acknowledgment}
The support of an EPSRC programme grant EP/L024926/1 is gratefully acknowledged by the authors.
M.B. acknowledges the financial support of Regione Autonoma della Sardegna (LR7 2010, grant `M4' CRP-27585). M.B. and A.B.M acknowledge the financial support of the European Community's Seven Framework Programme under contract number PIEF-GA-2011-302357-DYNAMETA.

\end{document}

%% file: macro.tex
\newcommand\eq[1] {(\ref{#1})}

\newcommand{\beqa}{\begin{eqnarray}}
\newcommand{\eeqa}[1]{\label{#1}\end{eqnarray}}
\newcommand{\bequ}{\begin{equation}}
\newcommand{\eequ}[1]{\label{#1}\end{equation}}

\newcommand{\Gt}{\theta}

\newcommand{\CF}{{\cal F}}

\newcommand{\CW}{{\cal W}}

\newcommand{\beq}{\begin{equation}}
\newcommand{\eeq}{\end{equation}}
\newcommand{\overliner}{\begin{eqnarray}}
\newcommand{\earr}{\end{eqnarray}}
\newcommand{\beqn}{\begin{equation*}}
\newcommand{\eeqn}{\end{equation*}}
\newcommand{\overlinern}{\begin{eqnarray*}}
\newcommand{\earrn}{\end{eqnarray*}}
\newcommand{\prt}{\partial}

\newcommand{\fr}{\frac}

\def\APL{ Appl. Phys. Letters}

\def\IP{ Inverse Probl.}

\def\JEM{ J. Eng. Math.}

\def\JMPS{ J. Mech. Phys. Solids}

\def\JOpt{ J. Opt.}

\def\MOM{ Mech. Materials}
\def\MRL{ Math. Res. Lett.}

\def\NJP{ New J. Phys.}

\def\PRB{ Phys. Rev. B}
\def\PRL{ Phys. Rev. Lett.}
\def\PRSLA{ Proc. R. Soc. Lond. A}

\def\RPP{ Rep. Prog. Phys.}